\documentclass[aip,jcp,amsmath,amssymb,reprint,floatfix,superscriptaddress]{revtex4-1}
\usepackage[utf8]{inputenc}
\usepackage[T1]{fontenc}
\usepackage[english]{babel}
\usepackage[dvipsnames,svgnames,x11names]{xcolor}
\usepackage{mathtools,dsfont,braket}
\usepackage{tensor}
\usepackage{nicefrac}
\usepackage[version=4]{mhchem}
\usepackage{graphicx}
\usepackage{nicefrac}
\usepackage{adjustbox}
\usepackage{float}
\usepackage{tabularx}
\usepackage{booktabs}
\usepackage{multirow}
\usepackage{longtable}
\usepackage{arydshln}
\usepackage{scrextend}
\usepackage{dcolumn}
\usepackage{bm}
\usepackage{microtype}
\usepackage{lipsum}
\usepackage{xr}

\begin{document}

\date{\protect{\today}}
\bibliographystyle{jcp}

\pagestyle{myheadings}
\markright{Unitary general coupled-cluster}
%%%%%%%%%%%%%%%%%%%%%%%%%%%%%%%%%%%%%%%%%%%%%%%%%%%%%%%%%%%%%%%%%%%%%%%
\def\w{\omega}
\def\ex{{\text{ex}}}
\def\nsep{{\text{nsep}}}
\def\sep{{\text{sep}}}
\def\eff{{\rm eff}}
\def\ao{{\rm ao}}
\def\RI{{\rm RI}}
\def\HF{{\rm HF}}
\def\SCF{{\rm SCF}}
\def\Rone{{\rm R1}}
\def\Rtwo{{\rm R2}}
\def\half{\frac{1}{2}}
\def\quart{\frac{1}{4}}
\def\bra#1{\langle #1|}
\def\ket#1{|#1\rangle}
\def\expec<#1|#2|#3>{\bra{#1}#2\ket{#3}}
\def\BRA#1{\Big\langle #1 \Big|}
\def\KET#1{\Big|#1\Big\rangle}
\def\EXPEC<#1|#2|#3>{\BRA{#1}#2\KET{#3}}
\def\Anti{\mathcal A}
\def\Symm{\mathcal S}
\def\Perm{\mathcal P}
\def\bft{{\bf t}}
\def\bfA{{\bf A}}
\def\bfB{{\bf B}}
\def\bfC{{\bf C}}
\def\bfF{{\bf F}}
\def\bfG{{\bf G}}
\def\unity{{\bf 1}}
\def\ms{{\small\it ms}}
\def\bfeta{\mbox{\boldmath$\eta$}}
\def\bfbeta{\mbox{\boldmath$\beta$}}
\def\bfalpha{\mbox{\boldmath$\alpha$}}
\def\bfzeta{\mbox{\boldmath$\zeta$}}
\def\bfxi{\mbox{\boldmath$\xi$}}
\def\bfgamma{\mbox{\boldmath$ \gamma $}}
\def\prop<<#1;#2>>{\langle\!\langle#1;#2\rangle\!\rangle}
\def\thalf{{\textstyle \frac{1}{2}}}
\def\tquart{{\textstyle \frac{1}{4}}}
\def\ddt#1{{\textstyle \frac{\partial #1}{\partial t}}}
\def\eps{\mbox{\large$\epsilon$}}
\def\e{\mbox{\large$e$}}
\def\rezi#1{{\textstyle\frac{1}{#1}}}
\def\mEh{m$E_\text{h}$}
\newcommand{\dtlde}[1]{\smash[t]{\tilde{\tilde{#1}}}}
\newcommand{\tldbr}[1]{\smash[t]{\tilde{\bar{#1}}}}
\newcommand{\brvbr}[1]{\smash[t]{\breve{\bar{#1}}}}
\newcommand{\ostar}[1]{\smash[t]{\overset{\star}{#1}}}

\newcommand{\rfn}[1]{\textsuperscript{\ref{#1}}}

\newcommand{\comment}[1]{{\color{blue}$\bullet$ #1 $\bullet$}}

\renewcommand{\topfraction}{0.9}	% max fraction of floats at top
\renewcommand{\bottomfraction}{0.8}	% max fraction of floats at bottom
 \setcounter{topnumber}{2}
 \setcounter{bottomnumber}{2}
 \setcounter{totalnumber}{4}     % 2 may work better
 \setcounter{dbltopnumber}{2}    % for 2-column pages
\renewcommand{\dbltopfraction}{0.9}
\renewcommand{\textfraction}{0.07}	
\renewcommand{\floatpagefraction}{0.7}	
\renewcommand{\dblfloatpagefraction}{0.7}	

\newif\iflongout
\longouttrue
%%%%%%%%%%%%%%%%%%%%%%%%%%%%%%%%%%%%%%%%%%%%%%%%%%%%%%%%%%%%%%%%%%%%%%%
\title{Capabilities and Limits of the Unitary Coupled-cluster Approach with Generalized Two-body Cluster Operators
       }
%%%%%%%%%%%%%%%%%%%%%%%%%%%%%%%%%%%%%%%%%%%%%%%%%%%%%%%%%%%%%%%%%%%%%%%
\author{ Andreas K\"ohn 
       }
\email{koehn@theochem.uni-stuttgart.de}
\affiliation{
  Institute for Theoretical Chemistry, University of Stuttgart, Pfaffenwaldring 55, 70569 Stuttgart
}
\author{ Jeppe Olsen
       }
\email{jeppe@chem.au.dk}
\affiliation{
  Department of Chemistry, University of {\AA}rhus,
  DK-8000 {\AA}rhus C, Denmark 
}

%%%%%%%%%%%%%%%%%%%%%%%% ABSTRACT %%%%%%%%%%%%%%%%%%%%%%%%%%%%%%%%%%%%%
\begin{abstract}
Unitary cluster expansions of the electronic wavefunction have recently gained much interest because of their use in conjunction with quantum algorithms. 
In this contribution, we investigate some aspects of an ansatz using generalized two-body excitations operators, which
has been considered in some recent works on quantum algorithms for quantum chemistry. 
Our numerical results show that in particular two-body operators with effective particle-hole excitation level of one in connection with the usual particle-hole double excitation operators lead to a very accurate yet compact representation of the wavefunction.  
Generalized two-body operators with effective excitation rank zero have a considerably less pronounced effect. 
We compare to standard and unitary coupled-cluster expansions and show that the above mentioned approach matches or even surpasses the accuracy of expansions with three-body particle-hole excitations, in particular at the onset of strong correlation.
A downside of the approach is that it is rather difficult to rigorously converge it to its variational minimum.
\end{abstract}
\maketitle
%\pagebreak

%%%%%%%%%%%%%%%%%%%%%%%%%%%%%%%%%%%%%%%%%%%%%%%%%%%%%%%%%%%%%%%%%%%%%%%
\section{Introduction}
%%%%%%%%%%%%%%%%%%%%%%%%%%%%%%%%%%%%%%%%%%%%%%%%%%%%%%%%%%%%%%%%%%%%%%%
\label{introduction}

Encoding the full configuration interaction (FCI) wavefunction by entangled quantum bits %(qbits) 
allows
for a logarithmically compact representation, which is one of the promising features of future applications of
quantum computers in the area of quantum chemistry and related fields.\cite{Bauer:CR120-12685,McArdle:RMP2020-015003} The realization of full eigensolvers, in particular
by quantum phase estimation algorithms,\cite{Aspuru-Guzik:Science2005-1704} is still hampered by available technology. For the realization of quantum chemical calculations on mid-term technology, so-called noisy intermediate scale quantum (NISQ) devices, hybrid algorithms like the variational quantum eigensolver (VQE) are preferred.\cite{Bharti:RMP2022-015004} For this, however, a compact representation of the variational ansatz for the wavefunction is required, for which the parameters are optimized. On the quantum side, the wavefunction may be expanded into a FCI space, which makes approaches attractive that beforehand were considered of purely academic use. This is the case for the now very intensely studied unitary coupled-cluster ansatz. It was  conceived as a theoretical concept early on,\cite{Kutzelnigg:JCP1982-3081,Kutzelnigg:TCA1991-349} but only implemented with its leading order terms, e.g. Ref.\ \onlinecite{Bartlett:CPL1989-133}, because of the otherwise unfavorable scaling even for the lowest-order expansion of the cluster operator, i.e. including double excitations. A first numerical study of the full UCC method was provided by Cooper and Knowles\cite{Cooper:JCP2010-234102} only a decade ago, see also the work of Evangelista.\cite{Evangelista:JCP2011-224102}

The method was then rediscovered in the context of quantum computing,\cite{Peruzzo:NComm2014-4213,Wecker:PRA2015-042303,Romero:QST2018-014008,Sokolov:JCP2020-124107}  and an analysis of its properties was recently given by Evangelista et al.\cite{Evangelista:JCP2019-244112}
The variant including up to two-electron clusters, unitary coupled-cluster with singles and doubles (UCCSD), however, has only limited accuracy implying the need for more accurate but still compact representations of the wavefunction.

A possible extension, without resorting to three-body clusters and higher, could be based on the observations of Nooijen\cite{Nooijen:PRL2000-2108} and Nakatsuji\cite{Nakatsuji:JCP2000-2949} which were inspired by early work of the latter
on the contracted Schr\"odinger equation.\cite{Nakatsuji:PRA1976-41} It was then conjectured, that the exponential of a generalized two-body operator was sufficient to map any determinant that has a non-vanishing overlap with an exact solution, to this exact solution of the Schr\"odinger equation. This conjecture was debated in some detail, including numerical studies\cite{Voorhis:JCP2001-5033,Piecuch:PRL2002-113001} and theoretical analysis.\cite{Davidson:PRL2003-123001,Ronen:PRL2003-123002,Mazziotti:PRA2004-012507,Mukherjee:CPL2004-174,Kutzelnigg:PRA2005-022502}
It was in particular discussed, whether the good performance seen for some examples was merely based on the so-called asymptotic solution, which corresponds to the imaginary time propagation of any initial state and that is known to project any trial wavefunction to the exact ground state. The parameterizations used at that time, however, were not unitary and it is in fact easy to see that the asymptotic solution is associated with the exponential of a symmetric (or more generally: Hermitian) operator. An antisymmetric (anti-Hermitian) operator would this not contain this asymptotic solution and could thus have been an interesting target for further tests.

Recently, Lee et al.\cite{Lee:JCTC2018-311} have revived the generalized two-body operators in the context of a unitary coupled-cluster wavefunction for VQE and showed a number of promising first results, albeit limited to 16 spin-orbitals in their numerical tests. They also introduced a $k$-fold sequence of individual UCC-like transformations, each restricted to only pair excitations from single (spatial) orbitals, called $k$-UpCCGSD (unitary pair coupled-cluster with generalized singles and doubles). Further studies based on this ansatz were provided by
Greene-Diniz and Muñoz Ramo.\cite{Greene-Diniz:IJQC2021-e26352}  Baumann et al\cite{Bauman:QST2021-034008} used this idea in conjunction with double unitary transforms to include further external correlation.
In a recent study, Rubin et al.\cite{Rubin:JCTC2022-1480} discussed further compression of the two-body operator by tensor decomposition in order to arrive at sufficiently shallow quantum algorithms.

This article is based on work pursued by the two authors approximately 17 years back and originally inspired by Nooijen's conjecture (more specifically: testing it by applying a unitary generalized two-body operator).
Being ignorant of all the possible developments in quantum computing, the method and its results were then deemed of little interest, but time has disproved our view. Even though many of the works mentioned above, in particular Refs.\ \onlinecite{Lee:JCTC2018-311,Greene-Diniz:IJQC2021-e26352}, have now already shown the main features of this approach, we
think that we still can provide some additional insight, which we would like to share here.
In particular, our numerical results hint at a further compactification of the parameterization, as we can show that the doubles amplitudes with excitation rank one give the most important contributions while other parts of the operator may be dropped. Furthermore, we add  benchmark results for a number of systems with large Hilbert spaces (up to order $10^6$ determinants resulting from up to 32 spin orbitals), which is important to better see the capabilities of the approach. Also, we provide a comparison to higher order traditional coupled-cluster and unitary coupled-cluster expansions, which further helps to assess the accuracy of the generalized unitary coupled-cluster approach.

%%%%%%%%%%%%%%%%%%%%%%%%%%%%%%%%%%%%%%%%%%%%%%%%%%%%%%%%%%%%%%%%%%%%%%%%
\section{The generalized unitary two-body parameterization}
%%%%%%%%%%%%%%%%%%%%%%%%%%%%%%%%%%%%%%%%%%%%%%%%%%%%%%%%%%%%%%%%%%%%%%%%

All examples in this work are based on the Born-Oppenheimer clamped-nuclei Hamiltonian
\begin{equation}
  \label{eq:ham}
  H = \sum_{pq} h^{q}_p \hat a^{p}_{q} + \frac{1}{4}\sum_{pqrs} g^{qs}_{pr} \hat a^{pr}_{qs} 
\end{equation}
with one- and two-electron integrals defined as
\begin{align}
  \label{eq:integrals}
  h^q_p &= \expec<\phi_p(1)|\half\Delta_1 + \hat v(1)|\phi_q(1)> \,,\\
  \begin{split}
  g^{qs}_{pr} &= \expec<\phi_p(1)\phi_r(2)|\frac{1}{r_{12}}|\phi_q(1)\phi_s(2)>\\&\qquad - \expec<\phi_p(1)\phi_r(2)|\frac{1}{r_{12}}|\phi_s(1)\phi_q(2)>
  \end{split}
\end{align}
where $\hat v$ is the operator of the Coulomb potential induced by the nuclei. The indices $p$, $q$, $r$, \ldots\ run over
all spin orbitals which are previously determined by, for instance, a Hartree-Fock calculation.

We express the correlated wavefunction as
\begin{align}
\label{eq:defPsi}
|\Psi_\text{UCC}\rangle = e^{\hat T - \hat T^\dagger}|\Phi_0 \rangle
\end{align}
where the reference function for the course of this work is a single reference determinant.
This defines occupied and unoccupied orbitals, indexed in the following by $i$, $j$, $k$, \ldots\ and $a$, $b$, $c$, \ldots, respectively. The standard unitary coupled cluster theory is based on pure particle-hole excitations with
\begin{align}
\label{eq:defTusual}
\hat T = \sum_{n=1}^{n_\text{max}} \hat T_n\,, \quad \hat T_n = \frac{1}{(n!)^2}\sum_{aibj\ldots} t^{ij\cdots}_{ab\cdots} \hat a^{ab\cdots}_{ij\cdots}
\end{align}
where the minimal choice is $n=2$ giving UCCSD. For larger choices of $n$ the number of amplitudes drastically increases.

 A generalized excitation operator with at most two-body excitations can be written as
\begin{equation}
  \label{eq:defG}
 \hat T_\text{GTB} = \hat T_{1,1} + \hat T_{1,0} + \hat T_{2,2} + \hat T_{2,1} + \hat T_{2,0}
\end{equation}
where the first subscript is the particle rank and the second subscript the excitation rank with respect to the reference determinant. The operators are defined as
\begin{align}
%\label{eq:}
\hat T_{1,1} &= \sum_{ai} t^i_a  \hat a^{a}_{i} \\
\hat T_{1,0} &= \sum_{ij} t^i_j  \hat a^{j}_{i} + \sum_{ab} t^b_a a^{a}_{b} \\
\hat T_{2,2} &= \frac{1}{4}\sum_{abij} t^{ij}_{ab}  \hat a^{ab}_{ij} \\
\hat T_{2,1} &= \frac{1}{2}\sum_{aijk} t^{ij}_{ak}  \hat a^{ak}_{ij} + \frac{1}{2}\sum_{abic} t^{ic}_{ab}  \hat a^{ab}_{ic}  \\
\hat T_{2,0} &= \frac{1}{4}\sum_{ijkl} t^{ij}_{kl}  \hat a^{kl}_{ij} + \frac{1}{4}\sum_{abcd} t^{cd}_{ab}  \hat a^{ab}_{cd} \,.
\end{align}
For the zero-excitation rank operators it is implicitly assumed that these are confined to their antisymmetric part, as by construction only this part contributes. The above definition contains some redundancies, e.g.
\begin{equation}
  \label{eq:singles_in_T}
  \sum_{r} \hat a^{ar}_{ir} = \sum_r (\hat a^{a}_{i}\hat a^{r}_{r} + \delta_{ri}\hat a^{a}_{i}) =
    \hat a^{a}_{i} ( \hat N + \delta_{ri} )  
\end{equation}
with $\hat N$ being the number operator. In the course of this work, we will therefore omit the single excitations. Note that we have shown in other work concerning multireference coupled-cluster theory that it is better to include the singles and to remove the redundant part from the higher-rank operator.\cite{Hanauer:JCP2011-204111,Koehn:MP2020-e1743889} This was mainly required to avoid size-consistency problems. This issue is for the current tests of minor interest and we will ignore it for sake of simplicity.

The wavefunction, eq.\ \eqref{eq:defPsi}, has the property of being normalized to unity, thus the energy expression is simply
\begin{align}
\label{eq:defEnergy}
 E = \langle \Phi_0 |e^{-\hat T + \hat T^\dagger} \hat H   e^{\hat T - \hat T^\dagger}|\Phi_0 \rangle
\end{align}
and the coefficients of $\hat T$ can be varied freely without violating the normalization condition. Thus, the most straight-forward approach is to optimize the wavefunction by requiring $\partial E / \partial t_\rho = 0$ (where $\rho$ stands for an arbitrary index tuple appearing in the definition of $\hat T$), see also next section.

\begin{figure}[t]
\begin{center}
\includegraphics[width=\columnwidth]{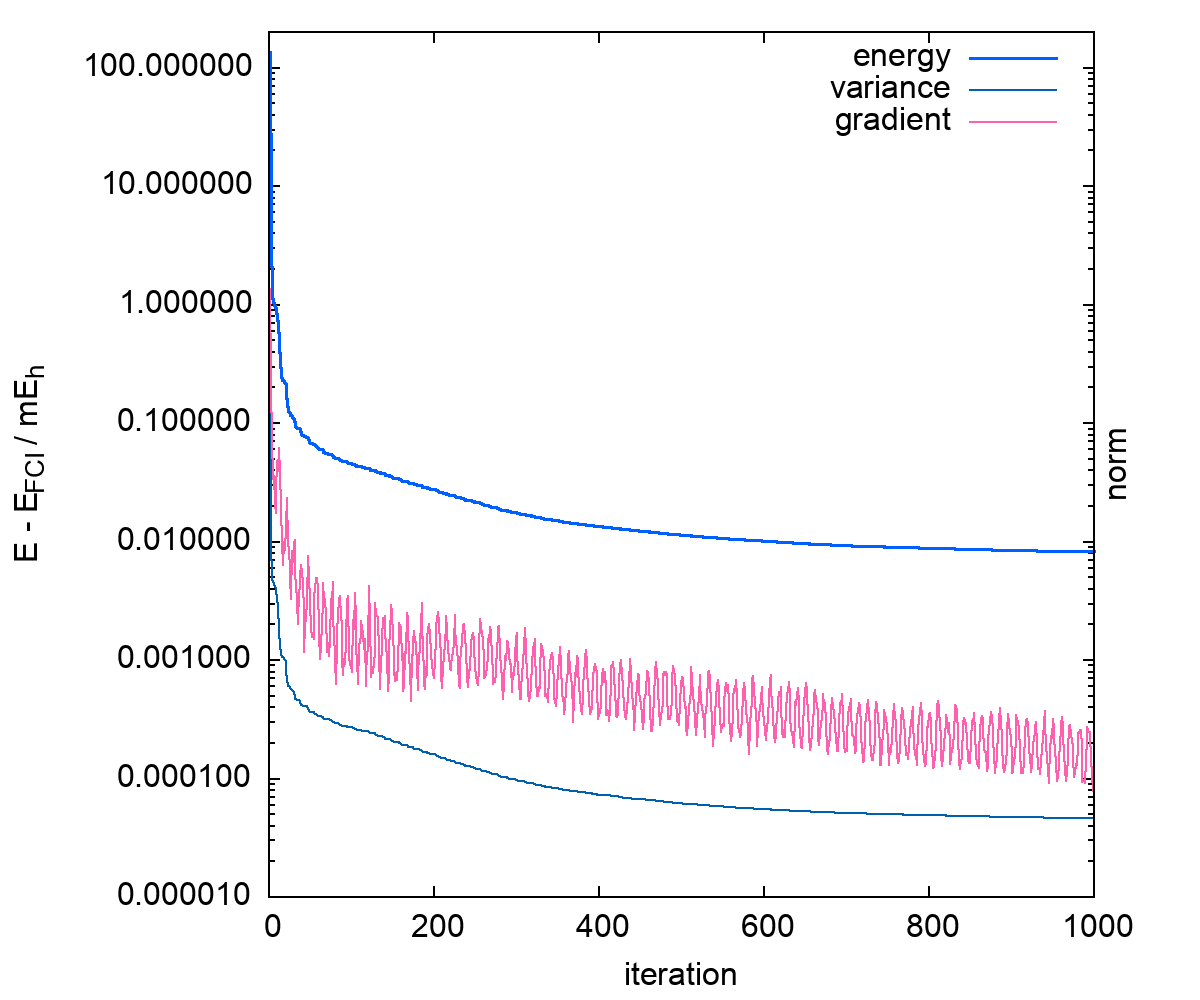}
\caption{Typical convergence trace of the tested generalized UCC methods (here: UCC(D$_1$,D$_2$)  for H$_2$O/VDZ at the equilibrium geometry).}
\label{fig:conv}
\end{center}
\end{figure}

%%%%%%%%%%%%%%%%%%%%%%%%%%%%%%%%%%%%%%%%%%%%%%%%%%%%%%%%%%%%%%%%%%%%%%%%
\section{Computational procedure}
%%%%%%%%%%%%%%%%%%%%%%%%%%%%%%%%%%%%%%%%%%%%%%%%%%%%%%%%%%%%%%%%%%%%%%%%
The computer code is based on the full configuration
interaction program LUCIA\cite{Olsen:CPL1990-169}, which allows for general manipulations
of CI expansions, see Ref.\ \onlinecite{Olsen:JCP2000-7140}.
 We use a series expansion approach to calculate
the action of $e^{\hat T - \hat T^\dagger}$ on a reference function, where the result is
projected onto a full-CI vector. The convergence is monitored via the
largest element of the current increment $1/n!(\hat T - \hat T^\dagger)^n\ket{\Psi_0}$.
Analytic first derivatives with respect to the parameters of $\hat T$ were
calculated by numerical integration of the Wilcox
identity\cite{Wilcox:JMP1967-962}, as described in Ref.\ 
\onlinecite{Voorhis:JCP2001-5033}, using a five-point Gauss quadrature.

Truncation errors are not an issue for the calculations
presented here: In general, all amplitudes remain well below 1.0 and
the series usually converges after 20-30 iterations with an residual
below $10^{-20}$, which is more than sufficient for double-precision
arithmetics. Numerical errors can be monitored through the norm of the
resulting wave-function which must be 1.0 as the wave-operator is
unitary. We only observed problems when we artificially set single
amplitudes to values above 30.0 or scaled $\hat T$ by overall factors of
this size.

To compare the  general unitary doubles parameterization with the usual
expansion with pure $n$-fold
 excitation operators in an unbiased way, we performed unitary
coupled cluster calculations in addition to standard projected coupled-cluster
calculations.
Unitary coupled-cluster was implemented along the same lines as
the generalized UCC, i.e.\ the exponential was
calculated as a series expansion acting on the reference, the result
again being a FCI vector. The gradient with respect to variations
of the amplitudes was again calculated using the Wilcox identity.

%%%%%%%%%%%%%%%%%%%%%%%%%%%%%%%%%%%%%%%%%%%%%%%%%%%%%%%%%%%%%%%%%%%%%%%%
\section{Results}
%%%%%%%%%%%%%%%%%%%%%%%%%%%%%%%%%%%%%%%%%%%%%%%%%%%%%%%%%%%%%%%%%%%%%%%%

While the standard and unitary coupled-cluster methods based on pure particle-hole excitations can be easily converged to gradient norms of $10^{-7}$ and better within 20 to 30 iterations, obtaining converged results for the generalized UCC approaches is much more difficult. This is a consequence of the structure of the ansatz, where certain operators annihilate the reference function but can give strong contributions in non-linear terms. Figure \ref{fig:conv} shows a typical computation using a simple conjugate gradient solver. The total energy is improved very rapidly within the first few iterations and after 50 iterations it is within 0.1 \mEh{} of the putative final result. After that, the rate of convergence dramatically decreases. We also computed the variance (giving another measure for the deviation from FCI), which also decreases monotonously during the computation. The gradient norm shows a notable oscillation, reflecting the complicated energy surface on which the optimum solution has to be located. Multiple minima on this surface may be an issue, as discussed by Lee et al.\cite{Lee:JCTC2018-311}, but we take confidence in their findings that multiple searches do not change conclusions about the final results. It is also our experience that changes in the details of the solver program do not have any significant effect on the energy after a few hundred iterations. Figure \ref{fig:conv} illustrates that the reported energies are a good estimate of the actual energy minimum that could be obtained with the investigated approach. It clearly is an upper bound and thus proves its accuracy. Any conclusions about restrictions of the method require a bit more care, as we cannot exclude incomplete convergence, but it seems unlikely that the final result will improve by more than one order of magnitude upon further optimization. It is of course also clear that this issue has to be solved in a practical way in order to turn this method into an applicable tool.

A first set of results can be found in Table \ref{tab:neon_water}. For Neon atom, all 8 valence electrons were correlated, and using a polarized valence double-$\zeta$ basis (cc-pVDZ), a FCI space of 3$\cdot$10$^{4}$ determinants is obtained. FCI computations with a slightly larger basis set (aug-cc-pVDZ) are also feasible (3$\cdot 10^{6}$ determinants) but become very costly for the evaluation of the generalized UCC methods as these require a much larger number of manipulations of the FCI vector and thus have not been carried out.
\begin{table*}[tb]
  \caption{Results for the systems Neon and water. $N_\text{coeff}$ is the formal number of freely variable coefficients (taking  into account spatial and spin-inversion symmetry). All energies (in \mEh{}) are reported as deviations from FCI (the reference result is given in the last line). For the generalized UCC computations, the gradient norm for the last iteration point is given in parentheses. VDZ denotes the cc-pVDZ basis without polarization functions. $R_e$ stands for the parameter set $R_\text{OH} = 95.785$ pm and $\alpha_\text{HOH} = 104.5^\circ$. $2 R_e$ stands for a doubled bond distance (symmetric stretch).}
  \label{tab:neon_water}
  \begin{adjustbox}{max width=\textwidth}
  \begin{tabular}{lrr@{}lrr@{}lr@{}lrr@{}lr@{}l}
  %\begin{tabular}{lSSSSSSSSSSSSS}
  \toprule
     &\multicolumn{3}{c}{Ne / cc-pVDZ}  
     &\multicolumn{3}{c}{H$_2$O / STO-6G ($R_e$) }   
     &\multicolumn{2}{c}{($2 R_e$)} 
     &\multicolumn{3}{c}{H$_2$O / VDZ ($R_e$) }
     &\multicolumn{2}{c}{($2 R_e$)} \\  
 \cmidrule(r){2-4}
 \cmidrule(r){5-7}
 \cmidrule(r){8-9}
 \cmidrule(r){10-12}
 \cmidrule(r){13-14}
 method & \multicolumn{1}{c}{$N_\text{coeff}$} & \multicolumn{2}{c}{$\Delta E$ / \mEh{}}  & 
          \multicolumn{1}{c}{$N_\text{coeff}$} & \multicolumn{2}{c}{$\Delta E$ / \mEh{}}  & \multicolumn{2}{c}{$\Delta E$ / \mEh{}}  & 
          \multicolumn{1}{c}{$N_\text{coeff}$} & \multicolumn{2}{c}{$\Delta E$ / \mEh{}}  & \multicolumn{2}{c}{$\Delta E$ / \mEh{}}  \\
\midrule     
CCSD    &   145 & 1.233 &  & 20 & 0.119 &  & 0.694 &  &   221 & 1.465 &  & 4.636 & \\
CCSDT   &  1163 & 0.160 &  & 32 & 0.024 &  & 0.473 &  &  1617 & 0.447 &  & 0.997 & \\
CCSDTQ  &  5780 & 0.010 &  & 40 & 0.000 &  & 0.000 &  &  6976 & 0.011 &  & 0.070 & \\
CCSDTQP & 15428 & 0.000 &  & 40 & 0.000 &  & 0.000 &  & 16944 & 0.003 &  & 0.010 & \\
UCCSD   &   145 & 1.025 &  & 20 & 0.101 &  & 0.242 &  &   221 & 1.042 &  & 3.639 & \\
UCCSDT  &  1163 & 0.009 &  & 32 & 0.002 &  & 0.030 &  &  1617 & 0.049 &  & 0.156 & \\
UCCSDTQ &  5780 & 0.001 &  & 40 & 0.000 &  & 0.000 &  &  6976 & 0.002 &  & 0.011 & \\
UCCSDTQP& 15428 & 0.000 &  & 40 & 0.000 &  & 0.000 &  & 16944 & 0.000 &  & 0.001 & \\
UCC(D$_1$,D$_2$)      &  783 &  0.014 & ($3\cdot 10^{-4}$) & 79 & 0.000 & ($4\cdot 10^{-5}$) & 0.000 & ($3\cdot 10^{-5}$)   & 1288 & 0.008 & ($1\cdot 10^{-4}$) & 0.019 & ($3\cdot 10^{-4}$) \\
UCC(D$_0$,D$_2$)      & 1351 &  1.090 & ($2\cdot 10^{-5}$) & 137 & 0.340 & ($7\cdot 10^{-6}$)  & 1.646 & ($3\cdot 10^{-4}$) & 2108 & 1.574 & ($1\cdot 10^{-4}$) & 7.097 & ($2\cdot 10^{-4}$) \\
UCC(S$_1$,D$_0$,D$_2$)      & 1357 &  1.019 & ($3\cdot 10^{-5}$) & 140 & 0.094 & ($1\cdot 10^{-4}$)  & 0.030 & ($6\cdot 10^{-4}$) & 2120 & 0.858 & ($2\cdot 10^{-4}$) & 0.902 & ($1\cdot 10^{-3}$) \\
UCC(D$_0$,D$_1$,D$_2$)& 1996 &  0.010 & ($1\cdot 10^{-4}$) & \footnotemark[1] & \footnotemark[1] &  & \footnotemark[1] &  & 3188 & 0.005 & ($3\cdot 10^{-4}$) & 0.007 & ($2\cdot 10^{-4}$) \\
FCI & 32523	& $-$128679.025	&&	40	&$-$75728.777	&&	$-$75644.555	&&	30968&	$-$76116.225 &&		$-$76028.440	\\
\bottomrule
  \end{tabular}
  \footnotetext[1]{Skipped as UCC(D$_1$D$_2$) is already overparameterized.}
  \end{adjustbox}
\end{table*}

In order assess the capabilities of the generalized UCC expansions, let us first review the convergence of the coupled-cluster hierarchy. Both traditional (projective) and unitary coupled-cluster computations were carried out, showing both an exponential convergence towards the FCI limit with excitation level. The UCC results are in line with values earlier reported by Evangelista\cite{Evangelista:JCP2011-224102} and display a slightly faster convergence than the projective coupled cluster results do. Turning to the generalized UCC, we find a clear picture, namely that the most important improvement over UCCSD comes from the $\hat T_{2,1}$ operators. Very good results are already obtained by only adding these operators, denoted as UCC(D$_1$,D$_2$) in Table \ref{tab:neon_water}. The result outperforms traditional CCSDT and is only slightly less accurate than UCCSDT (which is exceptionally good in this case, as will be seen in the other examples). The $\hat T_{2,0}$ operator does not have any significant effect, the result of UCC(D$_0$,D$_2$) is even less accurate than UCCSD due to the neglect of singles, which cannot be substituted by this operator. These are only contained in the $\hat T_{2,1}$ operator, see eq.\ \eqref{eq:singles_in_T}. Adding the pure single excitations $\hat T_{1,1}$ in the UCC(S$_1$,D$_0$,D$_2$) approach shows only little improvement, the result is more or less equivalent to UCCSD.  Adding both $\hat T_{2,1}$ and $\hat T_{2,0}$ operators brings the result slightly closer to the FCI limit, but in view of the strong increase of the number of coefficients, nearly three times as many are included for the UCC(D$_0$,D$_1$,$D_2$) ansatz as compared to UCC(D$_1$,D$_2$), the gain is rather limited. This leads us to our main work hypothesis: The UCC(D$_1$,D$_2$) ansatz is the optimal parametrization of a unitary coupled-cluster wavefunction, giving maximum accuracy with a minimal set of parameters. In particular, it should be a good alternative to UCCSDT.

Our next system is water in a minimal basis. We included this example, as it was reported in Ref.\ \onlinecite{Lee:JCTC2018-311} that the generalized UCC leads to the exact result in this case. Indeed, we can confirm this finding, both for the equilibrium structure and stretched water bonds, provoking the onset of static correlation effects. However, the simple reason for the good performance is that the FCI space is so small that even the UCC(D$_1$,D$_2$) ansatz is already overparameterized. On the other hand, the results for UCC(D$_0$,D$_2$) show, that a large number of parameters is not always sufficient: Even though it has more parameters than UCC(D$_1$,D$_2$), it does not provide an exact result (or at least: there is no straightforward way to locate a potentially exact solution with this parameterization). This  once more underlines that the $\hat T_{2,0}$ operators have only a small effect in the overall expansion.

More conclusive are probably the results for the larger case where the double zeta basis VDZ is used for the water molecule. The most notable outcome here is that UCC(D$_1$,D$_2$) for both the equilibrium and stretched geometry surpasses the accuracy of UCCSDT and nearly reaches full quadruples quality, despite a significantly smaller number of amplitudes in the ansatz. The UCC(D$_0$,D$_2$) approach clearly misses the effect of singles, as shown by the UCC(S$_1$,D$_0$,D$_2$) results, at the same time this also shows that the $\hat T_{2,1}$ operators in  UCC(D$_1$,D$_2$) effectively replace the singles.

\begin{table*}[tb]
  \caption{Results for the systems C$_2$, CH$_2$ and N$_2$.  All energies are given in \mEh{}, for further details see also Table \ref{tab:neon_water}. The C$_2$ bond distance is $R_e=127.0025$ pm, for CH$_2$ the geometry is given by $R_\text{CH} = 111.6563$ pm and $\alpha_\text{HCH} = 102.4^\circ$, for N$_2$ the bond distance is $R_e = 111.1272$ pm.}
  \label{tab:c2_ch2_n2}
  \begin{adjustbox}{max width=\textwidth}
  \begin{tabular}{lrr@{}lrr@{}lr@{}lrr@{}lr@{}l}
  %\begin{tabular}{lSSSSSSSSSSSSS}
  \toprule
     &\multicolumn{3}{c}{C$_2$ / VDZ}  
     &\multicolumn{3}{c}{CH$_2$  / cc-pVDZ ($R_e$)}   
     &\multicolumn{2}{c}{($2 R_e$)} 
     &\multicolumn{3}{c}{N$_2$  / VDZ ($R_e$)}
     &\multicolumn{2}{c}{($1.5 R_e$)} \\  
 \cmidrule(r){2-4}
 \cmidrule(r){5-7}
 \cmidrule(r){8-9}
 \cmidrule(r){10-12}
 \cmidrule(r){13-14}
 method & \multicolumn{1}{c}{$N_\text{coeff}$} & \multicolumn{2}{c}{$\Delta E$ / \mEh{}}  & 
          \multicolumn{1}{c}{$N_\text{coeff}$} & \multicolumn{2}{c}{$\Delta E$ / \mEh{}}  & \multicolumn{2}{c}{$\Delta E$ / \mEh{}}  & 
          \multicolumn{1}{c}{$N_\text{coeff}$} & \multicolumn{2}{c}{$\Delta E$ / \mEh{}}  & \multicolumn{2}{c}{$\Delta E$ / \mEh{}}  \\
\midrule     
CCSD      &    251 & 22.853 &  &    739 & 3.829 &  & 6.537 &  &    337 & 9.856 &  & 36.296 & \\
CCSDT     &   2731 &  1.873 &  &   9981 & 0.190 &  & 0.250 &  &   4369 & 1.950 &  &  9.333 & \\
CCSDTQ    &  18118 &  0.528 &  &  68612 & 0.006 &  & 0.013 &  &  35241 & 0.221 &  &  2.639 & \\ 
CCSDTQP   &  64734 &  0.129 &  & 231208 & 0.000 &  & 0.000 &  & 160293 & 0.026 &  &  0.912 & \\
CCSDTQPH  & 138196 &  0.007 &  & \footnotemark[1]&  &  &  &   & 448274 & 0.002 &  &  0.081 & \\
UCCSD     &    251 & 19.972 &  &    739 & 3.567 &  & 6.156 &  &    337 & 7.970 &  & 26.931 & \\
UCCSDT    &   2731 &  1.793 &  &   9981 & 0.122 &  & 0.318 &  &   4369 & 0.516 &  &  3.383 & \\
UCCSDTQ   &  18118 &  0.150 &  &  68612 & 0.002 &  & 0.009 &  &  35241 & 0.051 &  &  0.576 & \\
UCCSDTQP  &  64734 &  0.024 &  & 231208 & 0.000 &  & 0.000 &  & 160293 & 0.005 &  &  0.191 & \\
UCCSDTQPH & 138196 &  0.002 &  & \footnotemark[1] &  &  &  &  & 448274 & 0.000 &  &  0.018 & \\
UCC(D$_1$,D$_2$)      &   1786 &  0.269 & ($5\cdot 10^{-4})$ & 10348 & 0.019 & ($6\cdot 10^{-4}$) & 0.032 & ($3\cdot 10^{-4}$) & 1920 & 0.104 & ($5\cdot 10^{-5}$) & 0.763 & ($9\cdot 10^{-4}$)\\
UCC(D$_0$,D$_1$,D$_2$)   &   4934 &  0.043 & ($4\cdot 10^{-4})$ & 43438 & 0.004 & ($3\cdot 10^{-4}$) & 0.006 & ($2\cdot 10^{-4}$) & 4800 & 0.046 & ($5\cdot 10^{-4}$) & 0.259 & ($2\cdot 10^{-3}$)\\
FCI       & 208342 & $-$75649.857 &  & 394334 & $-$39023.280 &  & $-$38926.227 &  & 1196448 & $-$109116.473 &  & $-$108934.114 & \\
\bottomrule
  \end{tabular}
  \footnotetext[1]{Skipped as coupled-cluster expansions with 6-fold excitations are already equivalent to FCI for this example.}
  \end{adjustbox}
\end{table*}

The next set of tests, see Table \ref{tab:c2_ch2_n2}, focuses on cases with strong correlation effects. The carbon dimer, C$_2$, shows near-degeneracy of two configurations already at its equilibrium structure. This leads to very pronounced deviations of CCSD and UCCSD from the FCI limit and even for the expansions including three-electron clusters the errors are still large. Augmenting UCCSD with $\hat T_{2,1}$ operators, i.e.\ UCC(D$_1$,D$_2$), improves the energy greatly and  nearly matches UCCSDTQ quality. In this case, $\hat T_{2,0}$ seems to have a more significant impact, as it reduces the deviation to the FCI limit by a factor of five, see Table \ref{tab:c2_ch2_n2}.
This finding is confirmed by computations on singlet CH$_2$ and N$_2$ both also for stretched geometries. In the latter case this was shown for an underlying FCI expansion as large as $10^6$ determinants. We note, however, that for more strongly stretched bond distances of N$_2$ ($2 R_e$) we experienced convergence problems for all UCC variants, as the weight of the Hartree-Fock state diminishes, leading to large norms of the amplitudes. In this case, starting from a different reference state or using multiple consecutive expansions will be helpful, as suggested in Refs.\ \onlinecite{Lee:JCTC2018-311,Bauman:QST2021-034008}.

As a final example, we turn to twisted ethylene. Unfortunately, the extended run-times for the generalized unitary coupled-cluster approach only allow for treating this case with a minimal basis and frozen core orbitals. The most interesting issue about this system is that at 90$^\circ$ twist angle two degenerate configurations exist. It is thus an interesting (but rarely performed) test to compute energies beyond this point while staying on the now energetically disfavored reference function.
\begin{table*}[tb]
  \caption{Results for twisted ethylene (STO-6G minimal basis). Given are FCI energies (in $E_\text{h}$) and deviations to FCI (in \mEh{}). The structure parameters $R_\text{CH} = 109$ pm, $R_\text{CC} = 135$ pm, $\alpha_\text{HCH} = 117.2^\circ$ and D$_2$ point group symmetry were used, the twist angle between the CH$_2$ planes is varied. In all computations we strictly stay on the reference determinant that correlates to the one a the minimum geometry.}
  \label{tab:c2h4}
  \begin{adjustbox}{max width=\textwidth}
%  \begin{tabular}{lrrrrrrrrrrr}
  \begin{tabular}{ldddddddddd@{ }l}
  %\begin{tabular}{lSSSSSSSSSSSSS}
  \toprule
     &&&&\multicolumn{4}{c}{difference to FCI [\mEh{}]}  \\
 \cmidrule(r){3-12}
 angle & \multicolumn{1}{c}{$E_\text{FCI}$/ $E_\text{h}$} 
       & \multicolumn{1}{c}{CCSD} & \multicolumn{1}{c}{CCSDT} & \multicolumn{1}{c}{CCSDTQ} & \multicolumn{1}{c}{CCSDTQP}  
       & \multicolumn{1}{c}{UCCSD} & \multicolumn{1}{c}{UCCSDT} & \multicolumn{1}{c}{UCCSDTQ} & \multicolumn{1}{c}{UCCSDTQP}
       & \multicolumn{2}{c}{UCC(D$_1$,D$_2$)}    \\ 
\midrule     
  0 & -77.992275 & 1.281 &  0.472 &  0.008 & 0.002 & 0.863 & 0.106 & 0.002 & 0.000 & 0.089 &($9\cdot 10^{-5}$) \\
 10 & -77.990026 & 1.287 &  0.473 &  0.008 & 0.002 & 0.868 & 0.107 & 0.002 & 0.000 & 0.089 &($6\cdot 10^{-5}$) \\
 20 & -77.983326 & 1.304 &  0.476 &  0.008 & 0.003 & 0.884 & 0.110 & 0.002 & 0.000 & 0.088 &($2\cdot 10^{-4}$) \\
 30 & -77.972323 & 1.336 &  0.479 &  0.008 & 0.003 & 0.917 & 0.116 & 0.002 & 0.000 & 0.090 &($1\cdot 10^{-4}$) \\
 40 & -77.957320 & 1.391 &  0.478 &  0.009 & 0.003 & 0.980 & 0.125 & 0.002 & 0.000 & 0.096 &($2\cdot 10^{-4}$) \\
 50 & -77.938871 & 1.487 &  0.458 &  0.009 & 0.003 & 1.107 & 0.141 & 0.002 & 0.000 & 0.106 &($1\cdot 10^{-4}$) \\
 60 & -77.918021 & 1.664 &  0.377 &  0.009 & 0.003 & 1.381 & 0.168 & 0.003 & 0.000 & 0.125 &($7\cdot 10^{-5}$) \\
 70 & -77.896832 & 2.008 &  0.099 &  0.007 & 0.004 & 2.020 & 0.222 & 0.004 & 0.000 & 0.157 &($2\cdot 10^{-4}$) \\
 80 & -77.879401 & 2.586 & -0.851 &  0.001 & 0.007 & 3.513 & 0.339 & 0.006 & 0.000 & 0.215 &($2\cdot 10^{-4}$) \\
 85 & -77.874118 & 2.844 & -1.920 & -0.007 & 0.010 & 4.776 & 0.440 & 0.008 & 0.000 & 0.264 &($2\cdot 10^{-4}$) \\
 90 & -77.872245 & 2.811 & -3.612 & -0.018 & 0.019 & 6.366 & 0.578 & 0.011 & 0.001 & 0.336 &($3\cdot 10^{-4}$) \\
 95 & -77.874118 & 2.167 & -5.954 & -0.029 & 0.034 & 8.078 & 0.744 & 0.016 & 0.001 & 0.428 &($6\cdot 10^{-4}$) \\
100 & -77.879401 & 0.709 & -8.763 & -0.037 & 0.059 & 9.624 & 0.912 & 0.020 & 0.001 & 0.548 &($4\cdot 10^{-3}$) \\
\midrule
$N_\text{coeff}$ & \multicolumn{1}{c}{107334} &\multicolumn{1}{c}{252} & \multicolumn{1}{c}{2372} & \multicolumn{1}{c}{12510} & \multicolumn{1}{c}{37038} & \multicolumn{1}{c}{252} & \multicolumn{1}{c}{2372} & \multicolumn{1}{c}{12510} & \multicolumn{1}{c}{37038} & \multicolumn{1}{c}{1155} \\ 
\bottomrule
  \end{tabular}
  \end{adjustbox}
\end{table*}

In Table \ref{tab:c2h4} we have collected the main results. For the FCI wavefunction, we see that the energies beyond 90$^\circ$ twist angle match the values of their respective mirror images (e.g. the values for 80$^\circ$ and 100$^\circ$). For the cluster expansions, which inherently have a bias towards the chosen reference determinant, there are significant deviations, in particular for the low-order approximations. The effect is better comprehensible in a pictorial representation. 
\begin{figure}[h!]
\begin{center}
\includegraphics[width=\columnwidth]{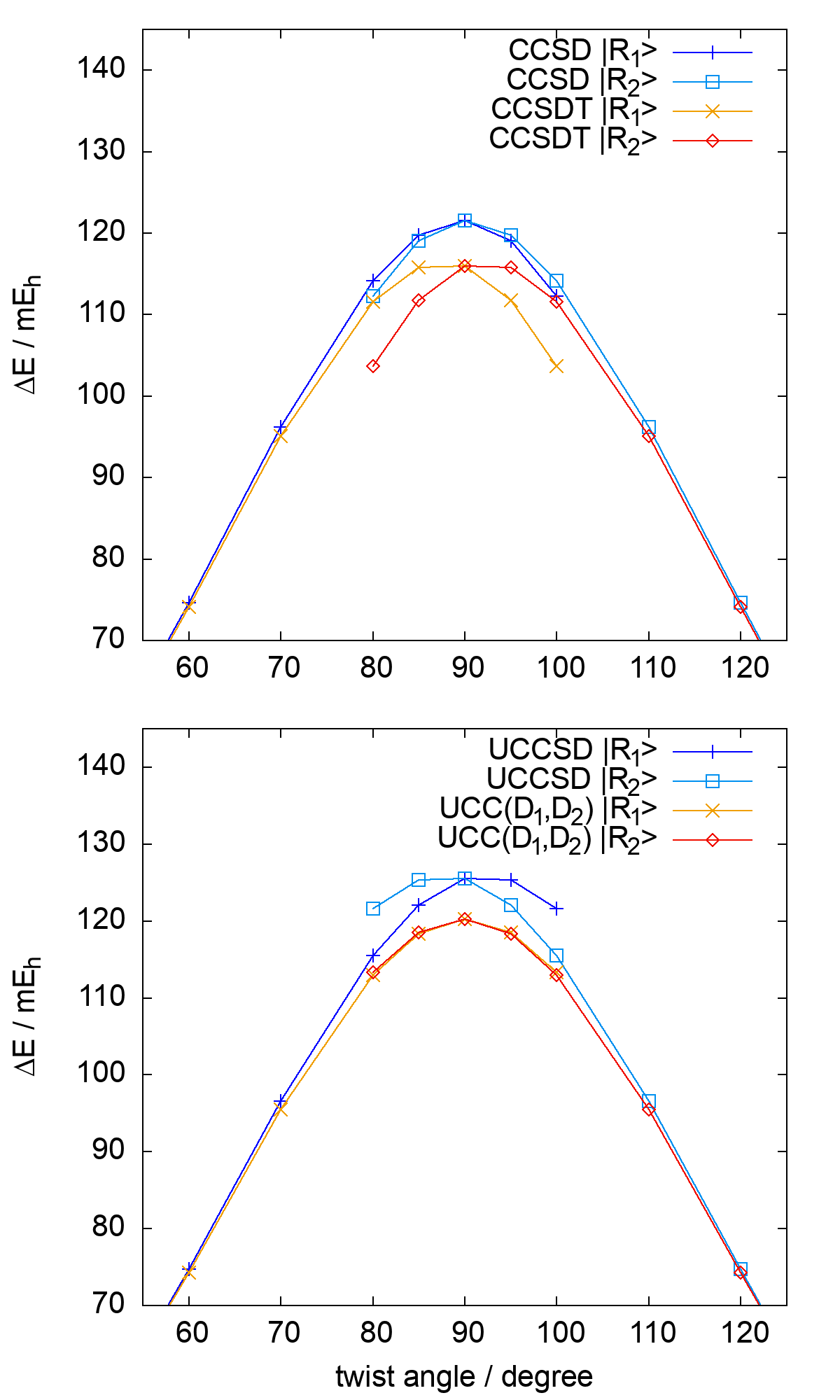}
\caption{Comparison of results with two different reference functions near to the fully twisted structure of ethylene. The energies are given relative to the values for zero twist angle for each approach.}
\label{fig:c2h4}
\end{center}
\end{figure}
In Figure \ref{tab:c2h4} the run of the curve around 90$^\circ$ is shown in comparison to the mirror image that corresponds to choosing the alternative reference determinant. CCSD fares astonishingly well for the investigated region, but this is partially based on error compensation and associated to the onset of overshooting the FCI limit. 
In fact, for CCSDT we observe partially strong overestimation of the total energy, as seen from the entries in Table \ref{tab:c2h4}. For the unitary variant, UCCSD, overshooting is prevented by the strict observation of the variation principle and the results reveal indeed the shortcomings of the cluster operator restricted to pure single and double excitations in this case. 
The generalized two-body expansion in the UCC(D$_1$,D$_2$) variant, on the other hand, works very well, giving errors consistently smaller than those of UCCSDT, although the number of amplitudes is only half as large (1155 compared to 2372). There still is a small difference between the results using either of the two reference determinants (the values for 80$^\circ$ and 110$^\circ$ differ by 0.3 \mEh{}), indicating that the approach is not invariant with respect to the choice of the reference state.

%%%%%%%%%%%%%%%%%%%%%%%%%%%%%%%%%%%%%%%%%%%%%%%%%%%%%%%%%%%%%%%%%%%%%%%%
\section{Conclusions}
%%%%%%%%%%%%%%%%%%%%%%%%%%%%%%%%%%%%%%%%%%%%%%%%%%%%%%%%%%%%%%%%%%%%%%%%

Our benchmark computations confirm the accuracy of a unitary coupled-cluster expansion with generalized two-body cluster operators. Most effective are two-body operators with excitation rank one, i.e. including the spin-orbital excitations $\hat a^{ab}_{ic}$ and $\hat a^{ak}_{ij}$ and denoted as $\hat T_{2,1}$ in this work. 
This may be understood by their property of generating approximations to higher-order clusters by non-linear terms, e.g.\ via $\hat T_3 \approx [\hat T_{2,1},\hat T_{2,2}]$, as indicated by the perturbational analysis of Kutzelnigg and Mukherjee.\cite{Kutzelnigg:PRA2005-022502} 
A similar approach has also been used to formulate explicitly correlated three-body clusters.\cite{Koehn:JCP2009-131101} 

A very compact and accurate ansatz for the unitary coupled-cluster wavefunction is obtained by augmenting the usual UCCSD wavefunction (with particle-hole excitations only) with $\hat T_{2,1}$. This approach is denoted as UCC(D$_1$,D$_2$) in this work. The $\hat T_{2,1}$ operator has on the order of $N_\text{occ}^2$ less amplitudes compared to the full three-body operator, where $N_\text{occ}$ is the number of occupied orbitals. Further compactification of the amplitudes may be possible using recent ideas for their decomposition.\cite{Rubin:JCTC2022-1480}

The comparison to the standard (particle-hole excitation-based) unitary coupled-cluster series confirms that the accuracy of UCC(D$_1$,D$_2$) matches that of UCCSDT and partly even surpasses it, in particular at the onset of strong correlation. In the latter case unitary coupled-cluster methods are also generally more stable than the standard projective approach.\cite{Cooper:JCP2010-234102}

The downside of the approach is that optimization with guarantees for sufficient convergence is challenging. However, if energies and gradients can be provided with sufficient speed and accuracy by a quantum algorithm, the approach will be an attractive improvement upon UCCSD.

\section*{Supplementary Material}

See supplementary material for a spreadsheet with all computed total energies.

%%%%%%%%%%%%%%%%%%%%%%%%%%%  ACKNOWLEDGMENT %%%%%%%%%%%%%%%%%%%%%%%%%% 
\begin{acknowledgments}
  This work is in part supported by the Deutsche Forschungsgemeinschaft (DFG, German Research Foundation)  
  by funding EXC 2075-390740016.
  A. K. also acknowledges the support by the Stuttgart Center for Simulation Science (SimTech).
\end{acknowledgments}

\section*{Author Declarations}

The authors have no conflicts to disclose.

\section*{Data availability}

The data that supports the findings of this study are available within the article and its supplementary material.

%%%%%%%%%%%%%%%%%%%%%%%%%%%%%%% APPENDIX %%%%%%%%%%%%%%%%%%%%%%%%%%%%%%
%\appendix
%\section{Appendix}

%%%%%%%%%%%%%%%%%%%%%%%%%%%%%% BIBLIOGRAPHY %%%%%%%%%%%%%%%%%%%%%%%%%%%%
\bibliography{gucc-lib}

\printfigures

%%%%%%%%%%%%%%%%%%%%%%%%%%%%%%%%%% END %%%%%%%%%%%%%%%%%%%%%%%%%%%%%%%%
\end{document}